\documentstyle[11pt]{article}
\begin{document}
\baselineskip=24pt plus 2pt
\vskip -5cm
\vskip 0.5cm

\begin{center}
{\large \bf Asymmetry Effects in Polarized Hadron Scattering}\\
\vspace{20mm}
Yaw-Hwang Chen, Su-Long Nyeo{\footnote{Author to whom all correspondence
should be addressed.\\
E-mail: t14269@MAIL.NCKU.EDU.TW.}}\\ 
and\\
Chung-Yi Wu\\ 
\vspace{5mm}

Department of Physics, National Cheng Kung University,\\
Tainan, Taiwan 701, R.O.C \\

\vspace{10mm}
\end{center}

\begin{center}
{\bf Abstract}
\end{center}

We calculate the single-spin and double-spin asymmetry differential cross 
sections for the polarized hadron scattering $PP\rightarrow l^+ l^- + jet$ 
up to $O(\alpha_s)$ by the helicity amplitude method.  Numerical results of 
the differential cross sections, which can be used to probe the spin 
contents of the proton, are obtained from several sets of polarized parton 
distribution functions.


\newpage
\noindent
{\large \bf I. Introduction}

In recent years, the determination of the spin contents of the nucleon 
has been the subject of a large number of work, which was initiated by the 
EMC experimental results [1] that the spin of the proton does not arise 
from its constituent quarks as expected in the na\"\i ve quark model; the 
valence quarks contribute very little to the spin of the nucleon.  

Presently, there are three possibilities for explaining the EMC 
and other subsequent experimental results [2,3,4],
namely (i) large sea quark polarization [5]; (ii) large gluon 
polarization [6]; and (iii) mediately large sea and gluon polarizations
[7].  Since a different explanation leads to a different set of 
polarized parton distribution functions, the determination of the 
polarized parton densities by other experiments like polarized 
hadron-hadron scattering is very important for understanding of the 
spin contents of the nucleon.  From the experiments, we can study 
the single-spin and double-spin asymmetries.  For both initial protons 
longitudinally polarized, we study the double-spin asymmetries.  
Here the inclusive cross sections for the incoming hadrons' longitudinal 
polarizations are either parallel or antiparallel.  From the 
experimental point of view, it is simpler to consider the single-spin 
asymmetries, where one of the initial hadron beams is longitudinally 
polarized and has positive or negative helicity.  A nonzero single-spin
asymmetry implies that some of the parton-parton scatterings involve 
parity-violating weak interactions.  Therefore, the single-spin 
asymmetries can also be used to probe parity-violating parton-parton 
subprocesses [8].

In this paper, we shall study the polarized $PP\rightarrow l^+ l^- + jet$ 
scattering by the helicity amplitude method [9].  Both the single-spin and 
double-spin asymmetries are studied.  At high energies, it is expected that
virtual photon and virtual $Z$ boson equally make important contributions 
to the lepton pair production.  We organize the paper as follows.  In 
section II, we shall use the helicity amplitude method to calculate the 
helicity amplitudes for the differential cross sections for the polarized 
hadron scattering.  In section III, we calculate numerically the single 
and double asymmetries for the scattering.  Finally, our conclusion is given 
in section IV.

\vskip 1cm
\noindent
{\large \bf II. Calculation of the Helicity Amplitudes}

In this section, we shall calculate the amplitudes for the single-spin 
and double-spin asymmetry differential cross sections for the polarized 
$PP \rightarrow l^+ l^- + jet$ scattering up to $O(\alpha_s)$ by the 
helicity amplitude method.  We need to consider the following two parton 
subprocesses and their charge conjugates contributing to the dilepton 
production
\begin{equation}
G + q \rightarrow l^+ l^- + q\,,
\end{equation}
\begin{equation}
q + \bar{q} \rightarrow l^+ l^- + G\,.
\end{equation}
We denote $M(\lambda_1, \lambda_2; \lambda_3, \lambda_4, \lambda_5)$
as the helicity amplitude, where $\lambda_1$ and $\lambda_2$ are the
helicities of the initial partons, while $\lambda_3$ $\lambda_4$ and 
$\lambda_5$ are the helicities of the lepton pair $l^+$, $l^-$, and the 
final parton, respectively.  The nonvanishing tree-level amplitudes read
\[ 
M(\lambda_1, \lambda_2; \lambda_3, \lambda_4, \lambda_5)
= \left\{\begin{array}{ll}
M(\lambda_1, \lambda_2; \lambda_3, -\lambda_3, \lambda_2),
&{\rm for}\,\,G+q \rightarrow l^+ l^- + q\,,\\
M(\lambda_1, -\lambda_1; \lambda_3, -\lambda_3, \lambda_5),
&{\rm for}\,\,q + \bar{q} \rightarrow l^+ l^- + G\,.
\end{array}\right.\]
The helicity amplitudes for figures 1.(a) and 1.(b) are 
$M(\lambda_1, \lambda_2; \lambda_3, -\lambda_3, \lambda_2)$, whereas 
those for figures 1.(c) and 1.(d) are 
$M(\lambda_1, -\lambda_1; \lambda_2, -\lambda_2, \lambda_3)$.

To evaluate the amplitudes, we first define some useful formulae and 
our convention.  We denote the positive and negative helicity states 
by $\mid A_\pm\rangle$, which have the following properties:
\begin{eqnarray}
(1\pm\gamma_5)\mid A_\mp \rangle &=& 0\,,\nonumber\\
\mid A_+\rangle^c &=& \mid A_- \rangle\,,\nonumber
\end{eqnarray}
\vskip -1.7cm
\begin{equation}
\end{equation}
\vskip -1.7cm
\begin{eqnarray}
\langle A_\mp \mid B_\pm \rangle &=&
   -\langle B_\mp \mid A_\pm \rangle\,,\nonumber\\
\langle A_+ \mid \gamma_\mu \mid B_{+} \rangle &=&
   \langle B_- \mid \gamma_\mu \mid A_- \rangle\,,\nonumber
\end{eqnarray}
and, by Fierz rearrangement theorem,
\begin{eqnarray}
& &\hskip -2cm\langle A_+ \mid \gamma_\mu \mid B_+\rangle
\langle C_- \mid \gamma^\mu \mid D_-\rangle
 =  2 \langle A_+ \mid D_- \rangle
        \langle C_- \mid B_+ \rangle\,,\nonumber\\
& & \nonumber
\end{eqnarray}
\vskip -2.5cm
\begin{equation}
\end{equation}
\vskip -2.5cm
\begin{eqnarray}
& & \nonumber\\
& &\hskip -1cm\langle A_- \mid B_+ \rangle
\langle C_- \mid D_+ \rangle
 =  \langle A_- \mid D_+ \rangle\langle C_- \mid
      B_+ \rangle
      +\langle A_- \mid C_+ \rangle\langle B_- \mid
      D_+ \rangle \,.\nonumber
\end{eqnarray}

The massless spinors with momentum $p$ and helicity $\lambda = \pm$ are
$u_\pm(p)$, $v_\pm(p)$, $\bar{u}_\pm(p)$ and $\bar{v}_\pm(p)$, which satisfy
\begin{eqnarray}
\not\!p u(p) &=& \not\!p v(p)\,\, = \,\,\bar{u}(p)\!\!\not\!p
                    \,\, =\,\,\bar{v}(p)\!\!\not\!p,\, p^2=0\,,\nonumber\\
(1\pm\gamma_5) v_\pm &=& (1\mp\gamma_5) u_\pm
                     \,\,=\,\,\bar{u}_\pm(1\pm\gamma_5)
                    \,\, =\,\,\bar{v}_\pm(1\mp\gamma_5)\,\, =\,\, 0\,,\\
\bar{u}_\pm(p) \gamma_\mu u_\pm(p) &=&
                    \bar{v}_\pm(p) \gamma_\mu v_\pm(p)
                     \,\,=\,\,2p_\mu\,.\nonumber
\end{eqnarray}
For massless momentum $p$, we can write
$ \not\!p = \mid p_+\rangle \langle p_+ \mid + \mid p_-\rangle
 \langle p_- \mid $ and use the convention
\begin{eqnarray}
u_\pm (p) &=& v_\mp(p)\,\, =\,\, \mid p_\pm \rangle\,,\nonumber\\
\bar{u}_\pm(p) &=& \bar{v}_\mp(p)\,\, =\,\, \langle p_\pm \mid\,\,,\\
\mid p_\mp \rangle & = & \mid p_\pm \rangle^c\,.\nonumber
\end{eqnarray}
For simplicity, we shall write
\begin{eqnarray}
\langle p_- \mid q_+ \rangle & = & \langle pq \rangle 
= -\langle qp \rangle \,,\nonumber\\
\langle q_+ \mid p_- \rangle & = & \langle pq \rangle^\ast
= -\langle qp \rangle^\ast\,,\\
\mid\langle pq \rangle\mid^2 & = & 2p\cdot q \,.\nonumber
\end{eqnarray}

Finally, the gluon helicities are chosen to be
\begin{eqnarray}
\not\!\epsilon^\pm_1 & = & {\pm \sqrt{2} \over \langle p^\mp_5 \mid
p^\pm_1 \rangle} [\mid p^\mp_1\rangle \langle p^\mp_5 \mid +
\mid p^\pm_5 \rangle \langle p^\pm_1 \mid ]~~{\rm for}~~G+q\rightarrow
l^+ l^- + q\,,\\
\not\!\epsilon^\pm_5 & = & {\pm \sqrt{2} \over \langle p^\mp_1 \mid
p^\pm_5 \rangle} [\mid p^\mp_5\rangle \langle p^\mp_1 \mid +
\mid p^\pm_1 \rangle \langle p^\pm_5 \mid ]~~{\rm for}~~q+\bar{q}
\rightarrow l^+ l^- + G\,.
\end{eqnarray}

At high energies, both the virtual photon and $Z$ boson contribute to the 
lepton pair production.  The helicity amplitudes are listed in Appendix A.
From the amplitudes, we can obtain the polarized parton differential cross 
sections as follows:
\begin{eqnarray}
d\hat{\sigma}^{(++)} \pm d\hat{\sigma}^{(+-)}& \propto 
     &(\mid M(++;+-+)\mid^2 +
       \mid M(++;-++)\mid^2\nonumber\\
 &   & \pm\mid M(+-;+--)\mid^2
       \pm\mid M(+-;-+-)\mid^2)_{\rm fig.1a,1b}\nonumber\\
 &   & \pm(\mid M(+-;+-+)\mid^2
       +\mid M(+-;-++)\mid^2\nonumber\\
 &   & +\mid M(+-;+--)\mid^2
       +\mid M(+-;-+-)\mid^2)_{\rm fig.1c,1d}\,,\nonumber
\end{eqnarray}
\begin{eqnarray}
d\hat{\sigma}^{(+)} \pm d\hat{\sigma}^{(-)}& \propto 
     &(\mid M(++;+-+)\mid^2 +
       \mid M(++;-++)\mid^2\nonumber\\
 &   & +\mid M(+-;+--)\mid^2
       +\mid M(+-;-+-)\mid^2\nonumber\\
 &   & \pm\mid M(-+;+-+)\mid^2
       \pm\mid M(-+;-++)\mid^2\nonumber\\
 &   & \pm\mid M(--;+--)\mid^2
       \pm\mid M(--;-+-)\mid^2)_{\rm fig.1a,1b}\nonumber\\
 &   & +(\mid M(+-;+-+)\mid^2
       +\mid M(+-;+--)\mid^2\nonumber\\
 &   & +\mid M(+-;-++)\mid^2
       +\mid M(+-;-+-)\mid^2\nonumber\\
 &   & \pm\mid M(-+;+--)\mid^2
       \pm\mid M(-+;+-+)\mid^2\nonumber\\
 &   & \pm\mid M(-+;-++)\mid^2
       \pm\mid M(-+;-+-)\mid^2)_{\rm fig.1c,1d}\,.\nonumber
\end{eqnarray}

To write the differential cross sections explicitly, we consider the 
center-of-mass frame of the incoming hadrons with momenta 
\begin{equation}
P_1 = {1\over 2} \sqrt{S} (1, 0, 0, 1)\,,\,\,
P_2 = {1\over 2} \sqrt{S} (1, 0, 0, -1)\,.\\
\end{equation}
Let the momentum of the virtual photon or $Z$ boson be $q$.  Then 
the Mandelstam invariants are:
\begin{eqnarray}
S & = & (P_1 + P_2)^2\,,\nonumber\\
T & = & (q - P_2)^2\,,\\
U & = & (q - P_1)^2\,.\nonumber
\end{eqnarray}
At the parton-level scattering processes, the parton momenta have
the explicit forms:
\begin{eqnarray}
p_1 & = & x_1 P_1\,,\nonumber\\
p_2 & = & x_2 P_2\,,\nonumber\\
p^\mu_5 & = & p^\mu_1 + p^\mu_2 -Q^\mu\,,\\
Q^\mu & = & p^\mu_3 + p^\mu_4\,,\nonumber\\
q^\mu & = & p^\mu_4 - p^\mu_3\,,\nonumber
\end{eqnarray}  
where $p_5$ is the momentum of the outgoing parton.
The Mandelstam invariants are:
\begin{eqnarray}
\hat{s} & = & (p_1 + p_2)^2 = x_1 x_2 S\,,\nonumber\\
\hat{t} & = & (p_5 - p_2)^2 = (Q - p_1)^2\,,\\
\hat{u} & = & (p_5 - p_1)^2 = (Q - p_2)^2\,.\nonumber
\end{eqnarray}

In the center-of-mass frame of incoming partons, the absolute squares of the 
{\it sum of the helicity amplitudes} for the subprocesses 
$G_{\lambda_1}(p_1) +$
$q_{\lambda_2}(p_2)$$\rightarrow$ $l^+_{\lambda_3}(p_3)$
$l^-_{\lambda_4}(p_4) +$ $q_{\lambda_5}(p_5)$ are (cf. Appendix B):
\begin{eqnarray}
\mid M(++;+-+) \mid^2_{\rm fig.1a,1b} &=& \left(  
        {2g^2 g_Z^4 R^2_q L^2_e
     \over[(q^2 - M^2_Z)^2 + M^2_Z\Gamma^2_Z]}
    + {8g^2 e^4 Q^2_q\over q^4}\right.\nonumber\\
&  &+ \left.{8g^2 g^2_Z e^2 Q_q R_q L_e(q^2 -M^2_Z) \over
       q^2[(q^2 - M^2_Z)^2 + M^2_Z\Gamma^2_Z]}\right)
    [{2\pi\over\hat{s}}\hat{t}^2\nonumber\\
&  & + {\pi\over\hat{s}}\hat{t}\hat{u}
     + \pi\hat{u} + 2\pi Q^2 + {\pi\over\hat{s}}\hat{u}^2\nonumber\\
&  & + {\pi\over3}{(Q^2\hat{t} - \hat{s}\hat{u})\over\hat{s}\hat{u}
      (\hat{s} - Q^2)^2}
      (-Q^2\hat{u}\hat{t} - Q^2 \hat{t}^2\nonumber\\
&  & +\hat{s}\hat{t}\hat{u} + Q^2 \hat{s}\hat{t} - \hat{s}^2 \hat{u}
      -\hat{s}\hat{u}^2)\nonumber\\
&  & + {4\pi\over3}{Q^2 (-\hat{s}\hat{t} 
     + \hat{u}\hat{t} + \hat{t}^2)
     \over (\hat{s} - Q^2 )^2 }]\,,\\
&  & \nonumber\\
\mid M(++;-++)\mid^2_{\rm fig.1a,1b}
&=& \mid M(++;+-+)\mid^2_{\rm fig.1a,1b}\,,\\
&  & \nonumber\\
\mid M(+-;+--)\mid^2_{\rm fig.1a,1b} &=& \left(
     {2g^2 g_Z^4 L^2_q L^2_e
     \over[(q^2 - M^2_Z)^2 + M^2_Z \Gamma^2_Z]}
    + {8g^2 e^4 Q^2_q\over q^4}\right.\nonumber\\
& & + \left.{8g^2 g^2_Z e^2 Q_q L_q L_e (q^2 -M^2_Z) \over
    q^2[(q^2 - M^2_Z)^2 + M^2_Z \Gamma^2_Z]}\right)\times\nonumber\\
& &[-{4\pi Q^2\over\hat{s}\hat{u}}({\hat{t}\over2} + {\hat{u}\over2}
      + Q^2 )^2\nonumber\\
& & - {\pi Q^2 \over 3\hat{s}\hat{u}(\hat{s} - Q^2 )^2}
      (Q^2 \hat{u} - \hat{s}\hat{t}\nonumber\\
& & + Q^2 \hat{t} - \hat{s}\hat{u})^2]\,,\\
& & \nonumber\\
\mid M(+-;-+-) \mid^2_{\rm fig.1a,1b}& 
\stackrel{R_e \leftrightarrow L_e}{=}& \mid M(+-;+--) 
\mid^{2}_{\rm fig.1a,1b}\,.
\end{eqnarray}
While for the subprocesses $q_{\lambda_1}(p_1) +
\bar{q}_{\lambda_2} (p_2)\rightarrow l^+_{\lambda_3}(p_3)
l^{-}_{\lambda_4}(p_4) + G_{\lambda_5}(p_5)$,
the absolute squares of the {\it sum of the helicity amplitudes} read
\begin{eqnarray}
\mid M(+-;+-+) \mid^2_{\rm fig.1c,1d} 
& =& \left({2g^2 g_Z^4 L^2_q L^2_e
     \over [(q^2 - M^2_Z)^2 + M^2_Z\Gamma^2_Z]}
    + {8g^2 e^4 Q^2_q \over q^4}\right.\nonumber\\
&  &+ \left.{8g^2 g^2_Z e^2 Q_q L_q L_e (q^2 -M^2_Z) \over
      q^2[(q^2 - M^2_Z)^2 + M^2_Z \Gamma^2_Z]}\right)\times\nonumber\\
&  & [\pi Q^2{\hat{t}\over\hat{u}}
    + {\pi\over3}{Q^2 (Q^2\hat{u} - \hat{s}\hat{t})^2 \over \hat{u}
     \hat{t}(\hat{s} - Q^2 )^2}\nonumber\\
&  & + {4\pi\over3}{Q^4\hat{s}\over
     (\hat{s} - Q^2 )^2}]\,,\\
& & \nonumber\\
\mid M(+-;-++)\mid^2_{\rm fig.1c,1d}
&\stackrel{R_e \leftrightarrow L_e}{=}&\mid M(+-;+-+) 
\mid^2_{\rm fig.1c,1d}\,,\\
& & \nonumber\\
\mid M(+-;+--) \mid^2_{\rm fig.1c,1d} & = & \left( 
    {2g^2 g_Z^4 L^2_q L^2_e
     \over[(q^2 - M^2_Z)^2 + M^2_Z\Gamma^2_Z]}
    + {8g^2 e^4 Q^2_q\over q^4}\right.\nonumber\\
& & +\left. {8g^2 g^2_Z e^2 Q_q L_q L_e (q^2 - M^2_Z)\over
       q^2 [ (q^2 - M^2_Z)^2 + M^2_Z\Gamma^2_Z]}\right)\times\nonumber\\
& & [-2\pi\hat{u} - 2\pi Q^2 -\pi{\hat{s}\hat{u}\over\hat{t}}
     - 2\pi Q^2{\hat{s}\over \hat{t}}\nonumber\\
& & + {2\pi\over 3}
     {(Q^2\hat{t} - \hat{s}\hat{u})^2\over\hat{u}(\hat{s} - Q^2)^2}
    +{\pi\over3}{\hat{s}(Q^2\hat{t} - \hat{u}\hat{s})^2\over
     \hat{u} \hat{t}(\hat{s} - Q^2)^2}\nonumber\\
& & + {4\pi\over3}{Q^2\hat{s}\hat{t}\over(\hat{s} - Q^2)^2}
     +{4\pi\over 3} {Q^2 \hat{s}^2 \over (\hat{s} - Q^2 )^2} 
     ]\,,\\
& & \nonumber\\
\mid M(+-;-+-) \mid^2_{\rm fig.1c,1d}
&\stackrel{R_e \leftrightarrow L_e}{=}& \mid M(+-;+--) 
\mid^{2}_{\rm fig.1c,1d}\,.
\end{eqnarray}

\vskip 1cm
\noindent
{\large \bf III.  Numerical Results}

In this section, the differential cross sections for the polarized 
processes will be calculated numerically.  For the double polarized 
hadron process, we have
\begin{equation}
{d\Delta\sigma \over dQ^2 dy dq^2_T} = \sum_{ij} \int dx_1 dx_2
  \Delta f_i (x_1) \Delta f_j (x_2) [ \hat{s} {d\Delta
  \hat{\sigma}_{ij}\over dQ^2 d\hat{t} d\hat{u}}]\,,
\end{equation}
where $y$ is the rapidity, 
\begin{equation}
\Delta f_i (x) = f^{(+)}_i (x) - f^{(-)}_i (x)\,,
\end{equation}
is the polarized structure function, and
\begin{equation}
d\Delta\hat{\sigma}_{ij} = {1\over2}(d\hat{\sigma}^{(++)}_{ij} -
d\hat{\sigma}^{(+-)}_{ij})\,.
\end{equation}
For the single polarized hadron process, we have
\begin{equation}
{d\Delta\sigma \over dQ^2 dy dq^2_T} = \sum_{ij} \int dx_1 dx_2
  \Delta f_i (x_1) f_j (x_2) [\hat{s} {d\Delta
  \hat{\sigma}_{ij}\over dQ^2 d\hat{t} d\hat{u}}]\,,
\end{equation}
\begin{eqnarray}
d\Delta \hat{\sigma}_{ij}& = &d\hat{\sigma}^{(+)}_{ij} -
      d\hat{\sigma}^{(-)}_{ij}\nonumber\\
& = & {1\over2} (d\hat{\sigma}^{(++)}_{ij} + d\hat{\sigma}^{(+-)}_{ij}
- d\hat{\sigma}^{(-+)}_{ij} - d\hat{\sigma}^{(--)}_{ij})\,.
\end{eqnarray}

In the rest frame of $\gamma^\ast(Z^0)$ [10], the momenta of the lepton 
pair and outgoing parton are:
\begin{eqnarray}
p^\mu_3 & = & {1\over2}(E' - q' \cos\alpha , q'\sin\theta
  - q\sin\alpha \cos\beta \cos\theta - E' \cos\alpha \sin\theta,
  \nonumber\\
&  & - q \sin\alpha \sin\beta,
  q' \cos\theta - E' \cos\alpha \cos\theta
  + q \sin\alpha \cos\beta \sin\theta )\,,\nonumber\\
p^\mu_4 & = & {1\over2}(E' + q' \cos\alpha , q' \sin\theta
  + q\sin\alpha \cos\beta \cos\theta + E' \cos\alpha \sin\theta,\\
&  & q \sin\alpha \sin\beta,
  q' \cos\theta + E' \cos\alpha \cos\theta
  - q \sin\alpha \cos\beta \sin\theta )\,,\nonumber\\
p^\mu_5 & = & (q', -q' \sin\theta , 0, -q' \cos\theta)\,,\nonumber
\end{eqnarray}
where $E' = {{\hat{s} + q^2}\over 2\sqrt{\hat{s}}}$ and
$q' = {{\hat{s} - q^2}\over 2\sqrt{\hat{s}}}$.
The above momenta are in the form $x^\mu = (t, x, y, z)$.  The momenta of 
the incident partons thus define the $z$ axis and the direction of the 
$\gamma^\ast(Z^0)$ defines the $x$-$z$ plane.  The angles $\alpha$
and $\beta$ describe the decay of the $\gamma^\ast(Z^0)$ relative
to its axes.  The phase space integration becomes
\begin{eqnarray}
& &{1\over (2\pi)^5}{d^3 p_3\over 2p^0_3}
{d^3 p_4\over 2p^0_4}{d^3 p_5\over 2p^0_5}
\delta^4(p_1 + p_2 - q - p_5)\nonumber\\
& & {\hskip 3cm}= {1\over (2\pi)^5} {1\over 16} d\Omega \pi \delta(\hat{s}
 +\hat{t} +\hat{u} - Q^2)
 ({dQ^2 d\hat{t} d\hat{u}\over\hat{s}})\,.
\end{eqnarray}
To calculate the differential cross sections at the hadron level,
we write the (double polarized) differential cross section as [11]
\begin{eqnarray}
{d\Delta \sigma \over dQ^2 dy dq^2_T} & = &
\sum_{ij} \int dx_1 dx_2 \Delta f(x_1) \Delta f(x_2)
[\hat{s}{d\Delta \hat{\sigma}_{ij} \over dQ^2 d\hat{t}d\hat{u}}]\nonumber\\
& \equiv & \int dx_1 dx_2\delta(\hat{s} + \hat{t} + \hat{u} - Q^2)
 f(x_1 ,x_2)\nonumber\\
& \equiv &\int^1_{\sqrt{\tau_+} e^y}{dx_1 f(x_1 ,x^\ast_2 )
\over x_1 S - \sqrt{s} (Q^2 + q^2_T)^{1\over 2} e^y}\nonumber\\
&  & + \int^1_{\sqrt{\tau_+} e^{-y}}{dx_2 f(x^\ast_1 ,x_2)
\over x_2 S - \sqrt{s}(Q^2 + q^2_T)^{1\over2} e^{-y}}\,,
\end{eqnarray}
where
\begin{equation}
\sqrt{\tau_+} = \sqrt{{q^2_T\over S}} +\sqrt{ \tau +
 {q^2_T\over S}};~~\tau={Q^2\over S}\,\,,
\end{equation}
and
\begin{eqnarray}
x^\ast_1 & = & {x_2\sqrt{S}(Q^2 + q^2_T)^{1\over2}
   e^{-y} - Q^2 \over x_2 S -\sqrt{S} (Q^2 + q^2_T)^{1\over2}
   e^y}\,,\\
x^\ast_2 & = & {x_1\sqrt{S}(Q^2 + q^2_T)^{1\over2}
   e^y - Q^2 \over x_1 S -\sqrt{S}(Q^2 + q^2_T)^{1\over2}
   e^{-y}}\,.
\end{eqnarray}

Finally, we evaluate the single-spin and double-spin asymmetries defined 
respectively as:
\begin{eqnarray}
A_L    &=& {{d\sigma^{(+)} - d\sigma^{(-)}}\over
           {d\sigma^{(+)} + d\sigma^{(-)}}}\,,\\
A_{LL} &=& {{d\sigma^{(++)} - d\sigma^{(+-)}}\over
           {d\sigma^{(++)} + d\sigma^{(+-)}}}\,.
\end{eqnarray}

To evaluate $A_{LL}$, we need polarized parton distribution functions, 
which we shall use those given by Cheng, {\it et al.} (CLW) [12] and 
those by Gl\"uck, {\it et al.} (GRSV) [13].  To obtain $A_L$, we 
need to use polarized parton distribution functions together with the 
corresponding unpolarized ones.  Specifically, the CLW sets should be used 
with the MRS(A') sets [14], while the GRSV sets should be used with the 
unpolarized GRV sets [15].

In the CLW sets, two different cases of polarized parton distributions 
in the gauge-invariant factorization scheme are used: (I) sea 
polarization $\Delta s(x) \neq 0,$ gluon polarization $\Delta G(x) = 0;$ 
and (II) $\Delta s(x) \neq 0, \Delta G(x) \neq 0$,
where in both cases $\Delta u(x) = \Delta d(x) = \Delta s(x)$ is assumed.   
Here the spin-dependent Altarelli-Parisi equations are 
applied directly to the gauge-invariant parton spin distributions. 

While in the GRSV sets, polarized deep inelastic lepton nucleon scattering was 
considered up to the next-to-leading order QCD within the framework of the 
radiative parton model.  The structure functions were subject to two 
different sets of theoretical constraints related to two different views 
concerning the flavor $SU_f(3)$ symmetry properties of hyperon 
$\beta$-decays [16].  One set is called the standard (ST) scenario, while 
the other set is called the valence (VA) scenario.

Figure 2 shows the unpolarized differential cross sections being plotted 
with $q_T$ and $\sqrt{S} = 500 GeV$.  Three different dilepton masses of
$Q = 10GeV, Q = 50GeV$ and $Q = M_Z$ are used.  The MRS(A') and the GRV sets,
which are defined up to the next-to-leading order in QCD, give 
essentially the same results. 

Figures 3a--3c and 4a--4c show the predictions for the single-spin 
asymmetries and double-spin asymmetries, respectively.  From Fig. 3a-3c, 
we note that the two CLW sets generally give larger differences between the 
single-spin asymmetries $|A_L|$ than the two GRSV sets.  The differences 
between the asymmetries from the two GRSV sets are essentially constant over
the given $q_T$ range.  The asymmetries from both the CLW and GRSV sets are 
large when the dilepton mass is equal to $M_Z$, due to the fact that 
the effects of the $Z$ interference become important when the dilepton 
mass is equal to $M_Z$.  From $Q = 10GeV$ 
to $Q = M_Z$, the asymmetries change by two orders of magnitude. 

From Fig.~4a--4c, the double-spin asymmetries $|A_{LL}|$ are of the same 
order of magnitude in the given $q_T$ range and for the three $Q$ values.
The CLW sets give comparatively larger differences between the asymmetries in
the same $q_T$ range than the GRSV sets, which give essentially the same 
differences between the asymmetries. 

Finally, in Fig.~5a-5d, we give the contributions of the subprocesses 
$G + q \rightarrow l^+ l^- + q$ (Fig.~1a-1b) and 
$q + \bar{q} \rightarrow l^+ l^- + G$ (Fig.~1c-1d) to the single-spin 
and double-spin asymmetries at $Q = M_Z$ and $\sqrt{S} = 500 GeV$.  
Fig.~5a-5b are obtained from the CLW sets, while Fig.~5c-5d are obtained 
from the GRSV sets.  We observe that quark-gluon and quark-quark contributions 
to the single-spin and double-spin asymmetries behave differently with 
increasing $q_T$.  The contributions of the subprocesses to the asymmetries 
obtained from the CLW sets are more dispersed and are larger than those from 
the GRSV sets, while the single-spin or double-spin asymmetries of the 
subprocesses obtained from the two GRSV sets have similar behavior.  
In general, the contributions from quark-gluon (Fig.~1a-1b) vary more 
significantly than that from quarks (Fig~1c-1d) alone.

\vskip 1cm
\noindent
{\large \bf IV. Conclusion}

In this paper, we have made a numerical study of the single-spin and 
double-spin asymmetries for the $PP\rightarrow l^+l^- + jet$ process.  
In this study, we have used, for convenience, the helicity amplitude 
method to calculate the scattering amplitudes and  the differential cross 
sections for the asymmetries.  Both the asymmetries of the angular distribution 
of lepton pair in the large transverse momentum Drell-Yan process at RHIC 
energies were studied.  Polarized parton distributions from CLW [12] and from 
GRSV [13] were used.  Both the CLW and the GRSV sets were obtained to 
the next-to-leading order.  We have used the MRS(A') [14] and the GRV [15] 
sets of unpolarized distribution functions, which are also given to 
the next-to-leading order.  The unpolarized differential cross 
sections obtained from these sets are essentially the same.

We observe that the two CLW sets generally give larger differences between the 
single-spin asymmetries $|A_L|$ than the two GRSV sets, and the asymmetries 
generally increase as the dilepton mass approaches $M_Z$, due to the 
large effects of the $Z$ interference from the parity-violating $Z$-fermion 
coupling.  The differences between the asymmetries from the two GRSV sets are 
essentially constant and small.  From $Q = 10GeV$ to $Q = M_Z$, the asymmetries 
change by two orders of magnitude. 

However, the double-spin asymmetries $|A_{LL}|$ are of the same 
order of magnitude in the same $q_T$ range and for the three $Q$ values.
The CLW sets give comparatively larger differences between the asymmetries for 
the same $q_T$ range than the GRSV sets, which give essentially constant
differences between the asymmetries.  We note that the asymmetries are nearly 
constant for different $Q$ values.  

We have also calculated the contributions of the subprocesses 
$G + q \rightarrow l^+ l^- + q$ (Fig. 1a-1b) and 
$q + \bar{q} \rightarrow l^+ l^- + G$ (Fig. 1c-1d) to the single-spin 
and double-spin asymmetries at $Q = M_Z$ and $\sqrt{S} = 500 GeV$.  
The single-spin and double-spin asymmetries behave differently 
with increasing $q_T$.  The asymmetries of the subprocesses obtained 
from the CLW sets are more dispersed than from the GRSV sets, while 
those obtained from the two GRSV sets have similar behavior.

We observe that, at high energies, $Z$ boson contributes significantly 
to the polarized process and induces large parity-violating 
single-spin asymmetries.  The single-spin asymmetries become 
large as the dilepton mass approaches the $Z$ peak for both the CLW and GRSV sets.
The double-spin asymmetries are generally larger than the single-spin asymmetries 
for any dilepton mass.  Therefore, measurements of the single-spin asymmetries 
may be used to pin down the spin structure of the proton for distinguishing the
two CLW sets.  However, the two GRSV sets give about the same results over 
large values of $q_T$ for different dilepton masses.

Finally, we should mention that Leader and Srihdar [17] have also made a 
detailed study on the polarized $PP\rightarrow l^+l^- + jet$ and obtained 
some similar results.  Their work and ours were based on the $O(\alpha_s)$
tree Feynman diagrams on the parton level.  One-loop radiative QCD corrections 
to the polarized parton subprocesses remain to be calculated and are 
difficult.  We note that, for the unpolarized Drell-Yan process, 
inclusion of the $O(\alpha_s^2)$ corrections changes the cross section 
at fixed target energies up to $10\%$ [18]. 

\vskip 1cm
\begin{center}
{\large \bf Acknowledgments}
\noindent
\end{center}

This research was supported by the National Science Council of the 
Republic of China under Contract Nos. NSC 86-2112-M-006-002 and 
NSC 86-2112-M-006-005.

\vskip 1cm
\begin{center}
{\large \bf Appendix A}
\noindent
\end{center}

Here we list the nonvanishing helicity amplitudes for the polarized hadron 
subprocesses.  With the contribution of a $Z$ boson, the amplitudes read
\begin{eqnarray}
M^Z_1(+,+;+,-,+) & = & {\sqrt{2} ig T^a g^2_Z R_q L_e \over
     (q^2 - M^2_Z + iM_Z \Gamma_Z)}{\langle 13\rangle^\dagger
     \langle 42\rangle \over \langle 51 \rangle}\,,\nonumber\\
M^Z_2(+,+;+,-,+) & = & {\sqrt{2} ig T^a g^2_Z R_q L_e \over
     (q^2 - M^2_Z + iM_Z\Gamma_Z)}{\langle 52\rangle
     \langle 35\rangle^\dagger
     \langle 42\rangle\over\langle 51 \rangle\langle 12\rangle
     }\,,\nonumber\\
M^Z_1(+,+;-,+,+) & = & {\sqrt{2} ig T^a g^2_Z R_q R_e \over
     (q^2 - M^2_Z + iM_Z\Gamma_Z)}{\langle 41\rangle^\dagger
     \langle 32\rangle\over\langle 15 \rangle}\,,\nonumber\\
M^Z_2(+,+;-,+,+) & = & {\sqrt{2} ig T^a g^2_Z R_q R_e \over
     (q^2 - M^2_Z + iM_Z\Gamma_Z)}{\langle 52\rangle
     \langle 45\rangle^\dagger
     \langle 32\rangle \over \langle 12\rangle\langle 51\rangle
     }\,,\nonumber\\
M^Z_2(+,-;+,-,-) & = & {\sqrt{2} ig T^a g^2_Z R_q L_e \over
     (q^2 - M^2_Z + iM_Z\Gamma_Z)}{\langle 45\rangle^2
     \langle 43\rangle^\dagger \over \langle 12\rangle\langle 51\rangle}
     \,,\nonumber\\
M^Z_2(+,-;-,+,-) & = & {\sqrt{2} ig T^a g^2_Z L_q R_e \over
     (q^2 - M^2_Z + iM_Z\Gamma_Z)}{\langle 35\rangle^2
     \langle 34\rangle^\dagger \over \langle 12\rangle\langle 51\rangle}
     \,,\nonumber\\
M^Z_2(-,+;+,-,+) & = & {\sqrt{2} ig T^a g^2_Z R_q L_e \over
     (q^2 - M^2_Z + iM_Z\Gamma_Z)}{\langle 35\rangle^{\dagger 2}
     \langle 43\rangle \over \langle 12\rangle^\dagger\langle 15\rangle
     ^\dagger} \nonumber\,,\nonumber\\
M^Z_2(-,+;-,+,+) & = & {\sqrt{2} ig T^a g^2_Z R_q R_e \over
     (q^2 - M^2_Z + iM_Z\Gamma_Z)}{\langle 45\rangle^{\dagger 2}
     \langle 34\rangle \over \langle 12\rangle^\dagger \langle 15\rangle
     ^\dagger} \,,\nonumber\\
M^Z_1(-,-;+,-,-) & = & {\sqrt{2} ig T^a g^2_Z L_q L_e \over
     (q^2 - M^2_Z + iM_Z\Gamma_Z)}{\langle 14\rangle
     \langle 23\rangle^\dagger \over \langle 15 \rangle
     ^\dagger}\,,\nonumber\\
M^Z_2(-,-;+,-,-) & = & {\sqrt{2} ig T^a g^2_Z L_q L_e \over
     (q^2 - M^2_Z + iM_Z\Gamma_Z)}{\langle 25\rangle^\dagger
     \langle 54\rangle \langle 23\rangle^\dagger
     \over \langle 12\rangle^\dagger \langle 15 \rangle^\dagger}
     \,,\nonumber\\
M^Z_1(-,-;-,+,-) & = & {\sqrt{2} ig T^a g^2_Z L_q R_e \over
     (q^2 - M^2_Z + iM_Z\Gamma_Z)}{\langle 13\rangle
     \langle 42\rangle^\dagger \over \langle 51\rangle^\dagger}\,,
     \nonumber\\
M^Z_2(-,-;-,+,-) & = & {\sqrt{2} ig T^a g^2_Z L_q R_e \over
     (q^2 - M^2_Z + iM_Z\Gamma_Z)}{\langle 24\rangle^\dagger
     \langle 53\rangle \langle 25\rangle^\dagger
     \over \langle 12\rangle^\dagger \langle 15 \rangle^\dagger}\,,
     \nonumber\\
M^Z_4(+,-;+,-,+) & = & {\sqrt{2} ig T^a g^2_Z L_q L_e \over
     (q^2 - M^2_Z + iM_Z\Gamma_Z)}{\langle 14\rangle^2
     \langle 43\rangle^\dagger
     \over \langle 15\rangle\langle 25\rangle
     }\,,\nonumber\\
M^Z_3(+,-;+,-,-) & = & {\sqrt{2} ig T^a g^2_Z L_q L_e \over
     (q^2 - M^2_Z + iM_Z\Gamma_Z)}{\langle 23\rangle^\dagger
     \langle 54\rangle \over\langle 15 \rangle^\dagger }\,,\nonumber\\
M^Z_4(+,-;+,-,-) & = & {\sqrt{2} ig T^a g^2_Z L_q L_e \over
     (q^2 - M^2_Z + iM_Z\Gamma_Z)}{\langle 21\rangle^\dagger
     \langle 14\rangle \langle 23\rangle^\dagger
     \over \langle 25\rangle^\dagger \langle 51 \rangle^\dagger }\,,\nonumber\\
M^Z_4(+,-;-,+,+) & = & {\sqrt{2} ig T^a g^2_Z L_q R_e \over
     (q^2 - M^2_Z + iM_Z\Gamma_Z)}{\langle 13\rangle^2
     \langle 34\rangle^\dagger \over \langle 25\rangle\langle 15 \rangle}
     \,,\nonumber\\
M^Z_3(+,-;-,+,-) & = & {\sqrt{2} ig T^a g^2_Z L_q R_e \over
     (q^2 - M^2_Z + iM_Z\Gamma_Z)}{\langle 53\rangle
     \langle 24\rangle^\dagger \over \langle 15 \rangle^\dagger
     }\,,\nonumber\\
M^Z_4(+,-;-,+,-) & = & {\sqrt{2} ig T^a g^2_Z L_q R_e \over
     (q^2 - M^2_Z + iM_Z\Gamma_Z)}{\langle 13\rangle
     \langle 24\rangle^\dagger \langle 21\rangle^\dagger
     \over \langle 25\rangle^\dagger \langle 51
     \rangle^\dagger}\,,\nonumber\\
M^Z_3(+,-;+,-,+) & = & {\sqrt{2} ig T^a g^2_Z R_q L_e \over
     (q^2 - M^2_Z + iM_Z\Gamma_Z)}{\langle 42\rangle
     \langle 35\rangle^\dagger \over \langle 15 \rangle }\,,\nonumber\\
M^Z_4(-,+;+,-,+) & = & {\sqrt{2} ig T^a g^2_Z R_q L_e \over
     (q^2 - M^2_Z + iM_Z\Gamma_Z)}{\langle 42\rangle
     \langle 31\rangle^\dagger
     \langle 12\rangle \over \langle 25\rangle \langle 15 \rangle
     }\,,\nonumber\\
M^Z_4(-,+;+,-,-) & = & {\sqrt{2} ig T^a g^2_Z R_q L_e \over
     (q^2 - M^2_Z + iM_Z \Gamma_Z)}{\langle 13\rangle^{\dagger 2}
     \langle 43\rangle \over \langle 25\rangle^\dagger \langle 51 \rangle
     ^\dagger}\,,\nonumber\\
M^Z_3(-,+;-,+,+) & = & {\sqrt{2} ig T^a g^2_Z R_q R_e \over
     (q^2 - M^2_Z + iM_Z\Gamma_Z) } { \langle 32\rangle
     \langle 45\rangle^\dagger\over\langle 15 \rangle }\,,\nonumber\\
M^Z_4(-,+;-,+,+) & = & {\sqrt{2} ig T^a g^2_Z R_q R_e \over
     (q^2 - M^2_Z + iM_Z\Gamma_Z)}{\langle 12\rangle
     \langle 41\rangle^\dagger \langle 32\rangle
     \over \langle 25\rangle \langle 15 \rangle}\,,\nonumber\\
M^Z_4(-,+;-,+,-) & = & {\sqrt{2} ig T^a g^2_Z R_q R_e \over
     (q^2 - M^2_Z + iM_Z\Gamma_Z)}{\langle 14\rangle^{\dagger
     2} \langle 34\rangle
     \over \langle 25\rangle^\dagger \langle 51 \rangle^\dagger}\,.\nonumber\\
\end{eqnarray}
The parameters $L_q = 2\tau_3 - 2Q_q \sin^2\theta_W, R_q = -2Q_q \sin^2 
\theta_W$ are, respectively, the left-handed and right-handed coupling 
constants, of quark coupled with the $Z$ boson, where $\tau_3$ is an $SU(2)$ 
isospin quantum number, $Q_q$ is the charge of the quark, $\theta_W$ is the
Weinberg angle, and $T^a$ is the $SU(3)$ generator with $a$ being the 
color index.  

With the contribution of a virtual photon, we have
\begin{eqnarray}
M^\gamma_1(+,+;+,-,+) & = & {2\sqrt{2} ig e^2 Q_q T^a \over
     q^2 }{\langle 31\rangle^\dagger
     \langle 42\rangle \over \langle 15 \rangle}\,,\nonumber\\
M^\gamma_2(+,+;+,-,+) & = & {2\sqrt{2} ig e^2 Q_q T^a\over
     q^2}{\langle 52\rangle
     \langle 35\rangle^\dagger
     \langle 42\rangle \over \langle 51 \rangle \langle 12\rangle
     }\,,\nonumber\\
M^\gamma_1(+,+;-,+,+) & = & {2\sqrt{2} ig e^2 Q_q T^a\over
     q^2}{\langle 41\rangle^\dagger
     \langle 32\rangle \over \langle 15 \rangle }\,,\nonumber\\
M^\gamma_2(+,+;-,+,+) & = & {2\sqrt{2} ig e^2 Q_q T^a\over
     q^2}{\langle 52\rangle
     \langle 45\rangle^\dagger
     \langle 32\rangle \over \langle 12\rangle\langle 51 \rangle
     }\,,\nonumber\\
M^\gamma_2(+,-;+,-,-) & = & {2\sqrt{2} ig e^2 Q_q T^a\over
     q^2}{\langle 45\rangle^2
     \langle 43\rangle^\dagger \over \langle 12\rangle\langle 51 \rangle}
     \,,\nonumber\\
M^\gamma_2(+,-;-,+,-) & = & {2\sqrt{2} ig e^2 Q_q T^a\over
     q^2}{\langle 35\rangle^2
     \langle 34\rangle^\dagger \over \langle 12\rangle\langle 51 \rangle}
     \,,\nonumber\\
M^\gamma_2(-,+;+,-,+) & = & {2\sqrt{2} ig e^2 Q_q T^a\over
     q^2 } { \langle 35\rangle^{\dagger 2}
     \langle 43\rangle \over \langle 12\rangle^\dagger\langle 15 \rangle
     ^\dagger}\,,\nonumber\\
M^\gamma_2(-,+;-,+,+) & = & {2\sqrt{2} ig e^2 Q_q T^a\over
     q^2 } { \langle 45\rangle^{\dagger 2}
     \langle 34\rangle \over \langle 12\rangle^\dagger\langle 15 \rangle
     ^\dagger}\,,\nonumber\\
M^\gamma_1(-,-;+,-,-) & = & {2\sqrt{2} ig e^2 Q_q T^a\over
     q^2 } { \langle 14\rangle
     \langle 23\rangle^\dagger \over \langle 15 \rangle^\dagger}
     \,,\nonumber\\
M^\gamma_2(-,-;+,-,-) & = & {2\sqrt{2} ig e^2 Q_q T^a\over
     q^2}{\langle 25\rangle\dagger
     \langle 54\rangle \langle 23\rangle^\dagger
     \over \langle 12\rangle^\dagger \langle 15 \rangle^\dagger}
     \,,\nonumber\\
M^\gamma_1(-,-;-,+,-) & = & {2\sqrt{2} ig e^2 Q_q T^a\over
     q^2}{\langle 13\rangle
     \langle 24\rangle^\dagger \over \langle 15 \rangle
     ^\dagger}\,,\nonumber\\
M^\gamma_2(-,-;-,+,-) & = & {2\sqrt{2} ig e^2 Q_q T^a\over
     q^2}{\langle 24\rangle^\dagger
     \langle 53\rangle \langle 25\rangle^\dagger
     \over \langle 12\rangle^\dagger \langle 15 \rangle^\dagger}\,\,,\nonumber\\
M^\gamma_4(+,-;+,-,+) & = & {2\sqrt{2} ig e^2 Q_q T^a\over
     q^2}{\langle 14\rangle^2
     \langle 43\rangle^\dagger
     \over \langle 25 \rangle \langle 15\rangle
     }\,,\nonumber\\
M^\gamma_3(+,-;+,-,-) & = & {2\sqrt{2} ig e^2 Q_q T^a\over
     q^2}{\langle 54\rangle
     \langle 23\rangle^\dagger \over \langle 15 \rangle^\dagger
     }\,,\nonumber\\
M^\gamma_4(+,-;+,-,-) & = & {2\sqrt{2} ig e^2 Q_q T^a\over
     q^2}{\langle 14\rangle
     \langle 23\rangle^\dagger
     \langle 21\rangle^\dagger \over \langle 25\rangle^\dagger \langle
     51 \rangle^\dagger}\,,\nonumber\\
M^\gamma_4(+,-;-,+,+) & = & {2\sqrt{2} ig e^2 Q_q T^a\over
     q^2}{\langle 13\rangle^2
     \langle 34\rangle^\dagger \over \langle 15\rangle\langle 25 \rangle}
     \,,\nonumber\\
M^\gamma_3(+,-;-,+,-) & = & {2\sqrt{2} ig e^2 Q_q T^a\over
     q^2}{\langle 53\rangle
     \langle 24\rangle^\dagger \over \langle 15 \rangle^\dagger
     }\,,\nonumber\\
M^\gamma_4(+,-;-,+,-) & = & {2\sqrt{2} ig e^2 Q_q T^a\over
     q^2}{\langle 21\rangle^\dagger
     \langle 24\rangle^\dagger
     \langle 13\rangle \over \langle 25\rangle^\dagger\langle
     51 \rangle^\dagger}\,,\nonumber\\
M^\gamma_3(-,+;+,-,+) & = & {2\sqrt{2} ig e^2 Q_q T^a\over
     q^2}{\langle 42\rangle
     \langle 35\rangle^\dagger \over \langle 15\rangle
     }\,,\nonumber\\
M^\gamma_4(-,+;+,-,+) & = & {2\sqrt{2} ig e^2 Q_q T^a\over
     q^2}{\langle 12\rangle \langle 42\rangle
     \langle 31\rangle^\dagger \over \langle 15\rangle\langle 25\rangle
     }\,,\nonumber\\
M^\gamma_4(-,+;+,-,-) & = & {2\sqrt{2} ig e^2 Q_q T^a\over
     q^2}{\langle 13\rangle^{\dagger 2}
     \langle 43\rangle \over \langle 15\rangle^\dagger \langle 25\rangle
     ^\dagger}\,,\nonumber\\
M^\gamma_3(-,+;-,+,+) & = & {2\sqrt{2} ig e^2 Q_q T^a\over
     q^2}{\langle 32\rangle
     \langle 45\rangle^\dagger \over \langle 15 \rangle}\,,\nonumber\\
M^\gamma_4(-,+;-,+,+) & = & {2\sqrt{2} ig e^2 Q_q T^a\over
     q^2}{\langle 41\rangle^\dagger
     \langle 32\rangle \langle 12\rangle
     \over\langle 25\rangle \langle 15 \rangle}\,,\nonumber\\
M^\gamma_4(-,+;-,+,-) & = & {2\sqrt{2} ig e^2 Q_q T^a\over
     q^2}{\langle 14\rangle^{\dagger2}
     \langle 34\rangle
     \over \langle 25\rangle^\dagger\langle 51\rangle^\dagger}\,\,.\nonumber
\end{eqnarray}

\vskip 1cm
\begin{center}
{\large \bf Appendix B}
\noindent
\end{center}

In this appendix, we list the relations between the invariant variables 
$s_{ij}$ and the Mandelstam invariants.
\begin{eqnarray}
s_{12} & = & \hat{s}\nonumber\\
s_{13} & = & 2p_{1}\cdot p_{3}\nonumber\\
& = & {1\over 2} (Q^2 -\hat{t}) -{{Q^2 \hat{u} -\hat{s}\hat{t}}
\over 2(\hat{s} -Q^2 )} \cos \alpha - {\sqrt{Q^2 \hat{s}\hat{t}
\hat{u}}\over {\hat{s} -Q^2}} \sin \alpha \cos \beta \nonumber\\
s_{14} & = & 2p_{1}\cdot p_{4}\nonumber\\
& = & {1\over 2} (Q^2 -\hat{t}) +{{Q^2 \hat{u} -\hat{s}\hat{t}}
\over 2(\hat{s} -Q^2)} \cos \alpha + {\sqrt{Q^2 \hat{s}\hat{t}
\hat{u}}\over {\hat{s} -Q^2}} \sin \alpha \cos \beta \nonumber\\
s_{15} & = & 2p_{1}\cdot p_{5} = -\hat{u}\nonumber\\
s_{23} & = & 2p_{2}\cdot p_{3}\nonumber\\
& = & {1\over 2}(Q^2 -\hat{u}) -{{Q^2 \hat{t} -\hat{s}\hat{u}}\over
2(\hat{s} -Q^2 ) } \cos \alpha +{\sqrt{Q^2 \hat{s}\hat{t}\hat{u}}\over
{\hat{s} -Q^2}} \sin \alpha \cos \beta \nonumber\\
s_{24} & = & 2p_{2}\cdot p_{4}\nonumber\\
& = & {1\over 2} (Q^2 -\hat{u}) + {{Q^2 \hat{t} -\hat{s}\hat{u}}\over
2(\hat{s} -Q^2)} \cos \alpha -{\sqrt{Q^2 \hat{s}\hat{t}\hat{u}}\over
{\hat{s} -Q^2}} \sin \alpha \cos \beta \nonumber\\
s_{25} & = & 2p_{2}\cdot p_{5} = -\hat{t}\nonumber\\
s_{34} & = & 2p_{3}\cdot p_{4} = Q^2\nonumber\\
s_{35} & = & 2p_{3}\cdot p_{5}
 =  -{{\hat{u} + \hat{t}}\over 2}(1-\cos \alpha)\nonumber\\
s_{45} & = & 2p_{4}\cdot p_{5}
 =  -{{\hat{u} + \hat{t}}\over 2} (1+\cos \alpha)\,\,.\nonumber
\end{eqnarray}

\vskip 1cm

\begin{center}
{\large \bf FIGURE CAPTIONS}
\end{center}
\vskip 1cm
\noindent

{\large\bf Fig.~1}  Feynman diagrams for the $PP \rightarrow l^+ l^- + jet$.
     Curly lines denote gluons, wavy lines denote $\gamma^\ast$ or $Z$ 
     bosons and solid lines denote quarks and lepton pairs.

{\large\bf Fig.~2}  Unpolarized differential cross sections for the
    $PP\rightarrow l^+l^- + jet$ process at $\sqrt{S} = 500GeV$, as a 
    function of $q_T$, with $\tau$ is fixed at $Q^2 /S$ and $y=0$, at 
    various dilepton masses: $Q = 10 GeV, 50 GeV, M_Z$.  The MRS(A') set 
    and the GRV set give essentially the same results.

{\large\bf Fig.~3}  Single-spin asymmetries for the $PP\rightarrow 
    l^+ l^- + jet$ process at $\sqrt{S} = 500GeV$, as a function of $q_T$.
    The asymmetries shown in these figures are constructed out of
    the cross-sections differential in $q_T$, rapidity $y$ and $\tau$,
    with $\tau$ being fixed at $Q^2 /S$ and $y=0$.  The full and dotted lines 
    correspond respectively to the CLW Set I and Set II, which are 
    used together with the MRS(A') unpolarized distribution functions.  
    The circle-full and dotted-full lines correspond respectively to 
    the GRSV standard (ST) and valence (VA) scenario sets, which are used together 
    with the GRV sets.  Three figures corresponding to (a) $Q = 10GeV,$ 
    (b) $Q = 50GeV$ and (c) $Q = M_Z$ are shown. 

{\large\bf Fig.~4}  Double-spin asymmetries for the $PP\rightarrow l^+ l^- + 
    jet$ process at $\sqrt{S} = 500GeV$, as a function of $q_T$.
    Three figures corresponding to (a) $Q = 10GeV,$ (b) $Q = 50GeV$
    and (c) $Q = M_Z$ are shown.

{\large\bf Fig.~5}  Contributions of the subprocesses 
    $G + q \rightarrow l^+ l^- + q$ (Fig.~1a-1b) and 
    $q + \bar{q} \rightarrow l^+ l^- + G$ (Fig.~1c-1d) to the single-spin 
    and double-spin asymmetries at $Q = M_Z$ and $\sqrt{S} = 500 GeV$.
    (a) and (b) are contributions obtained from the CLW sets, while 
    (c) and (d) are those obtained from the GRSV sets.  

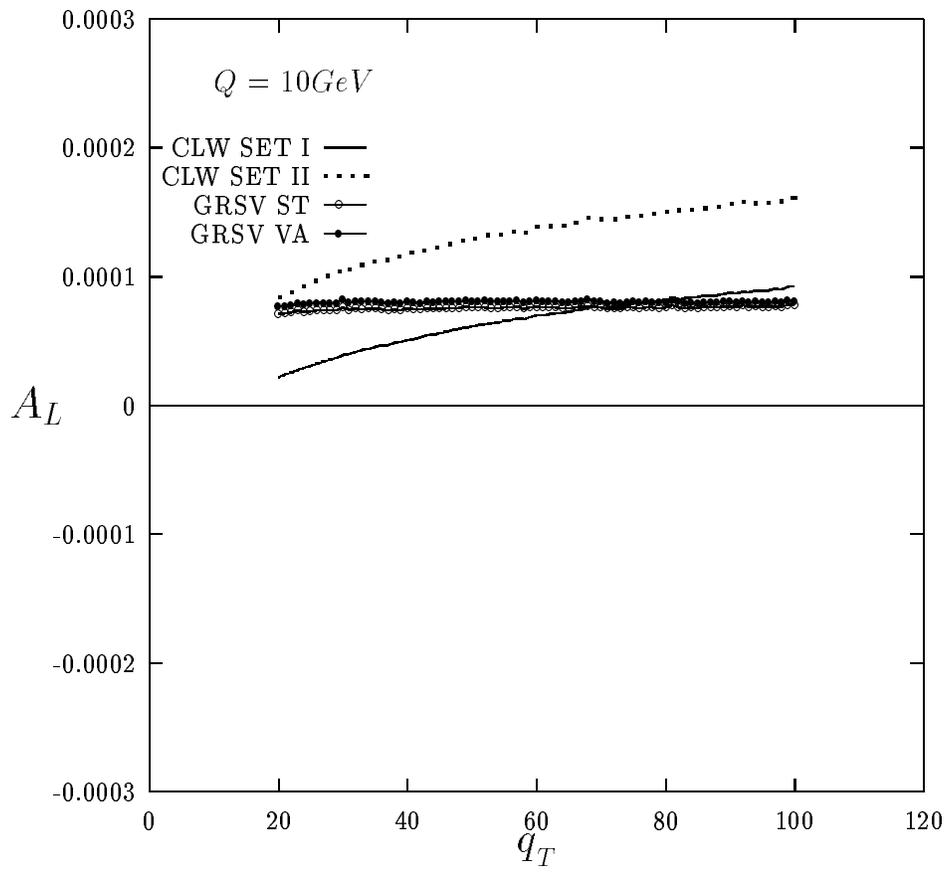
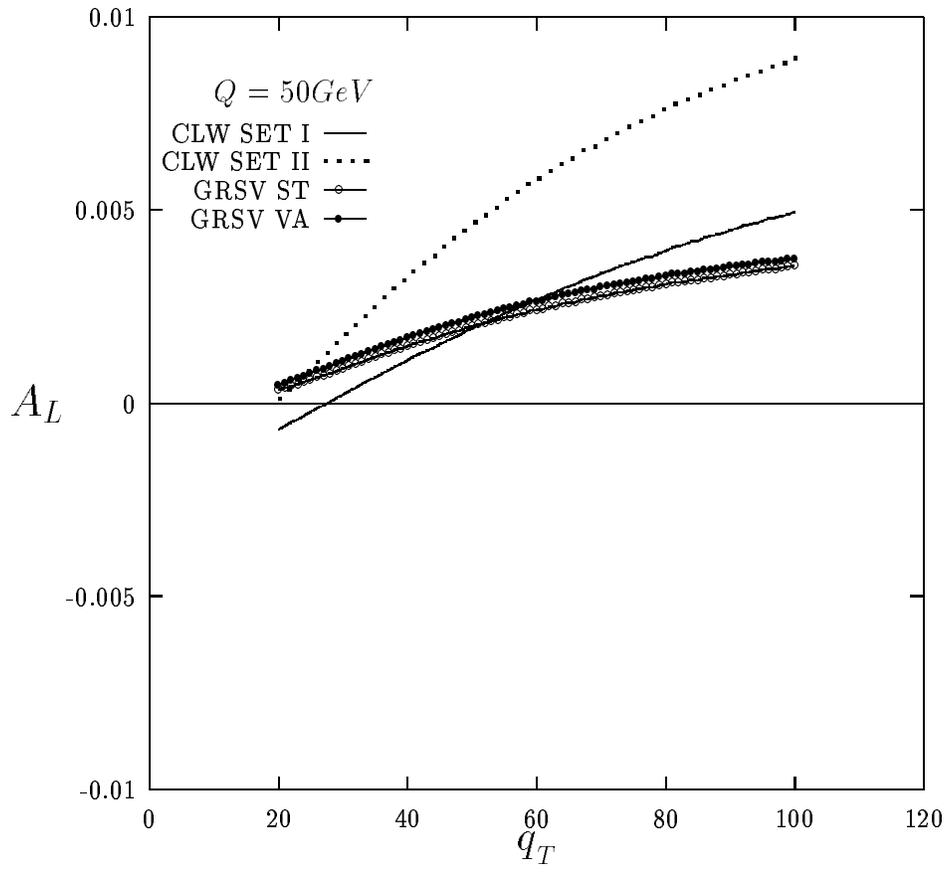
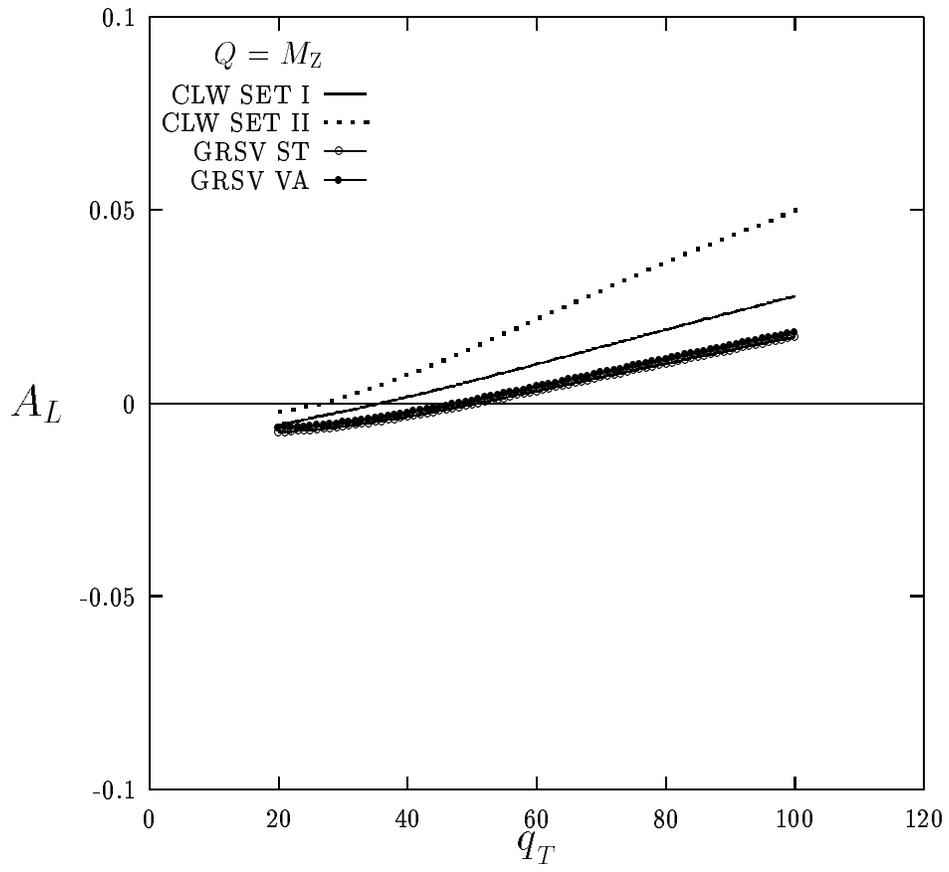
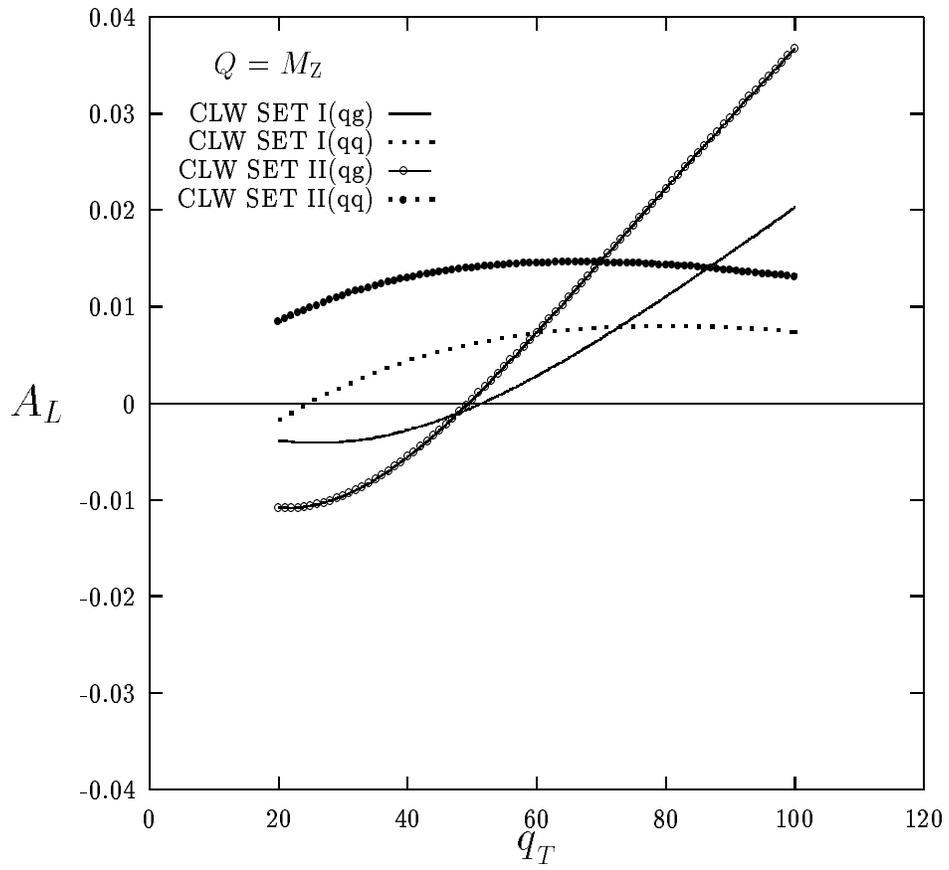
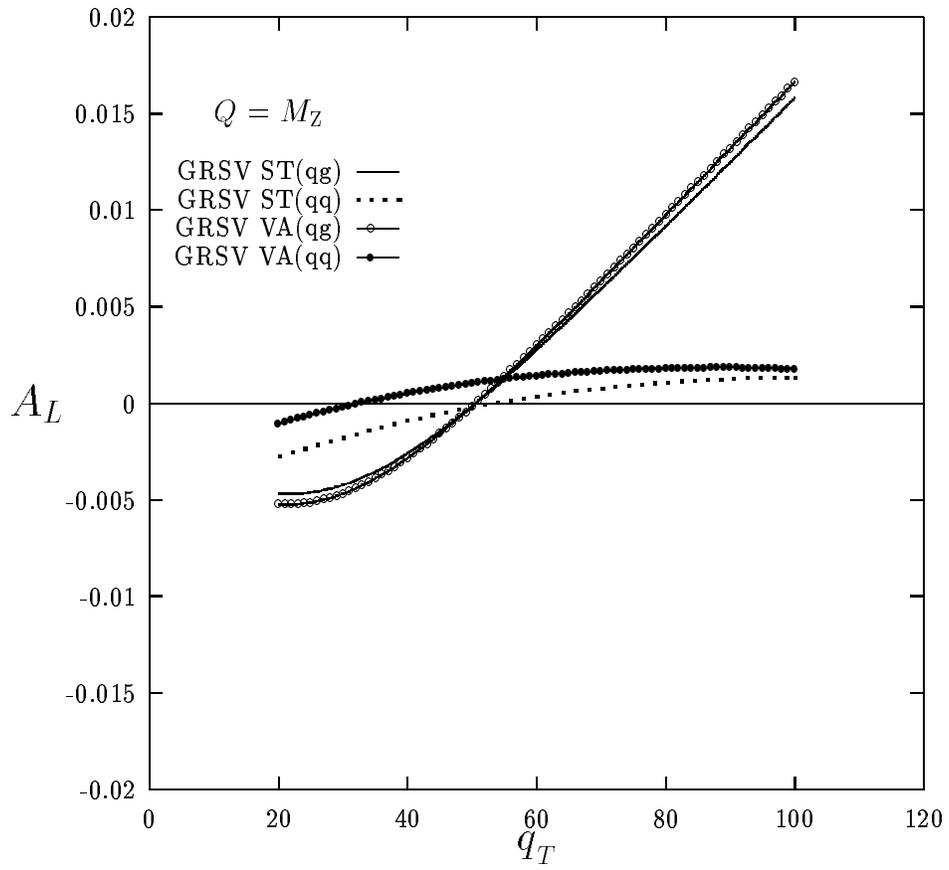
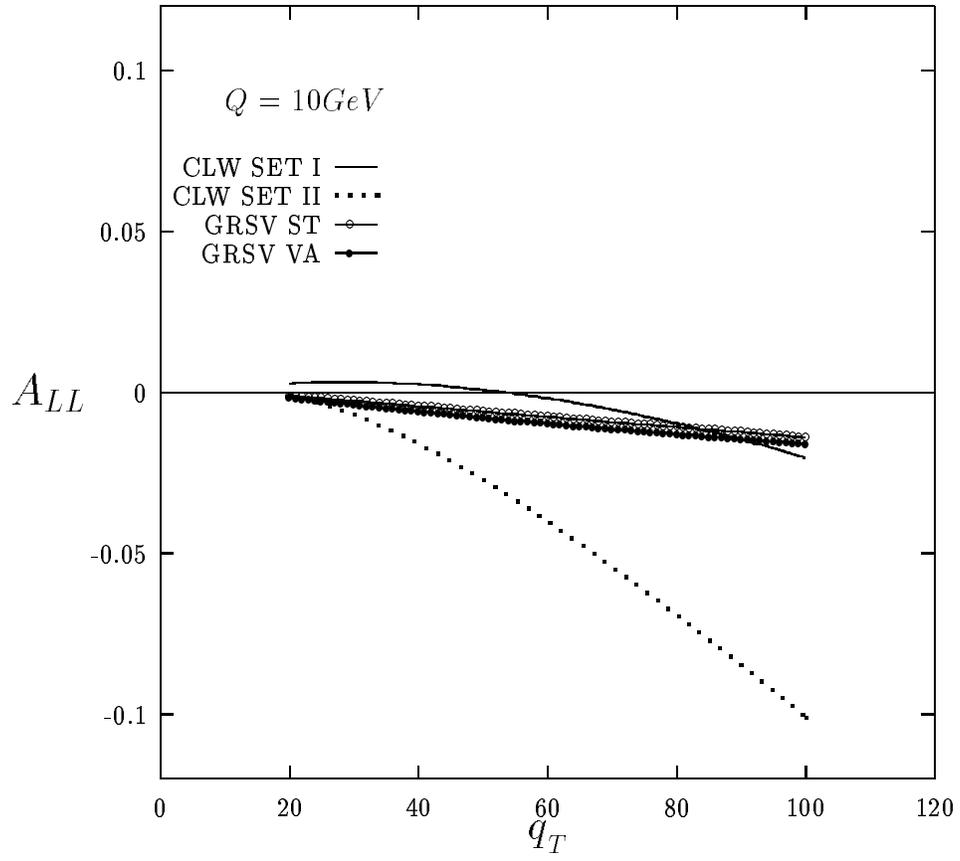
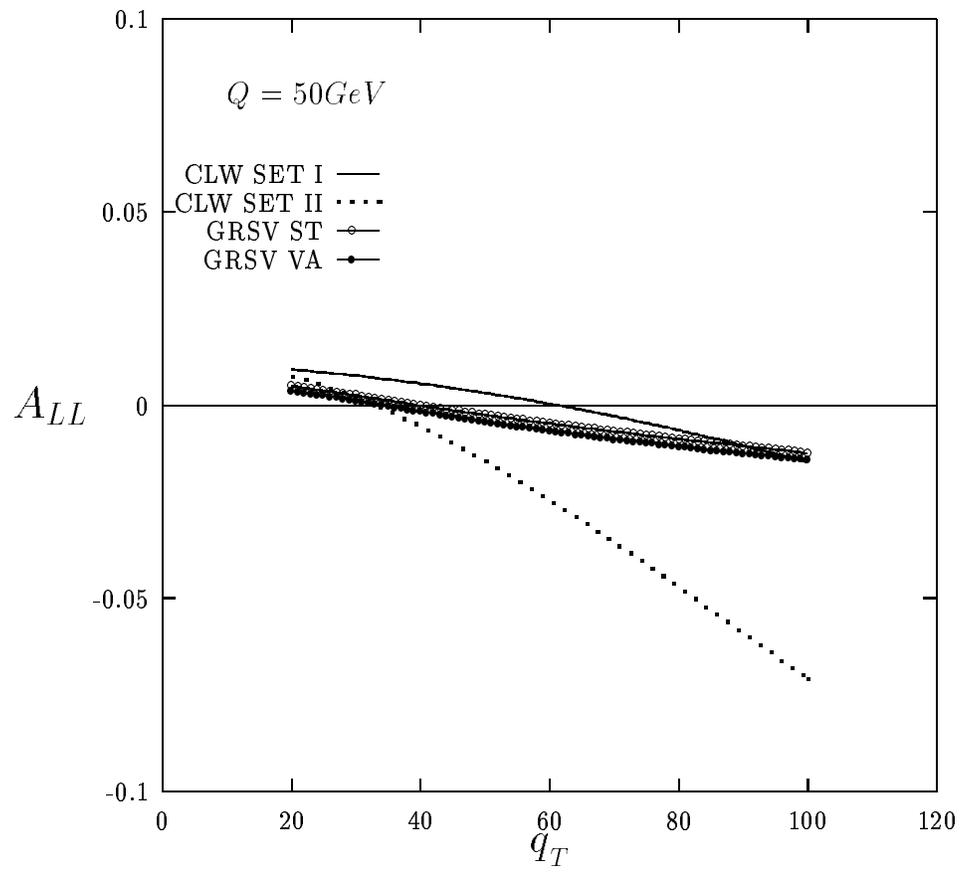
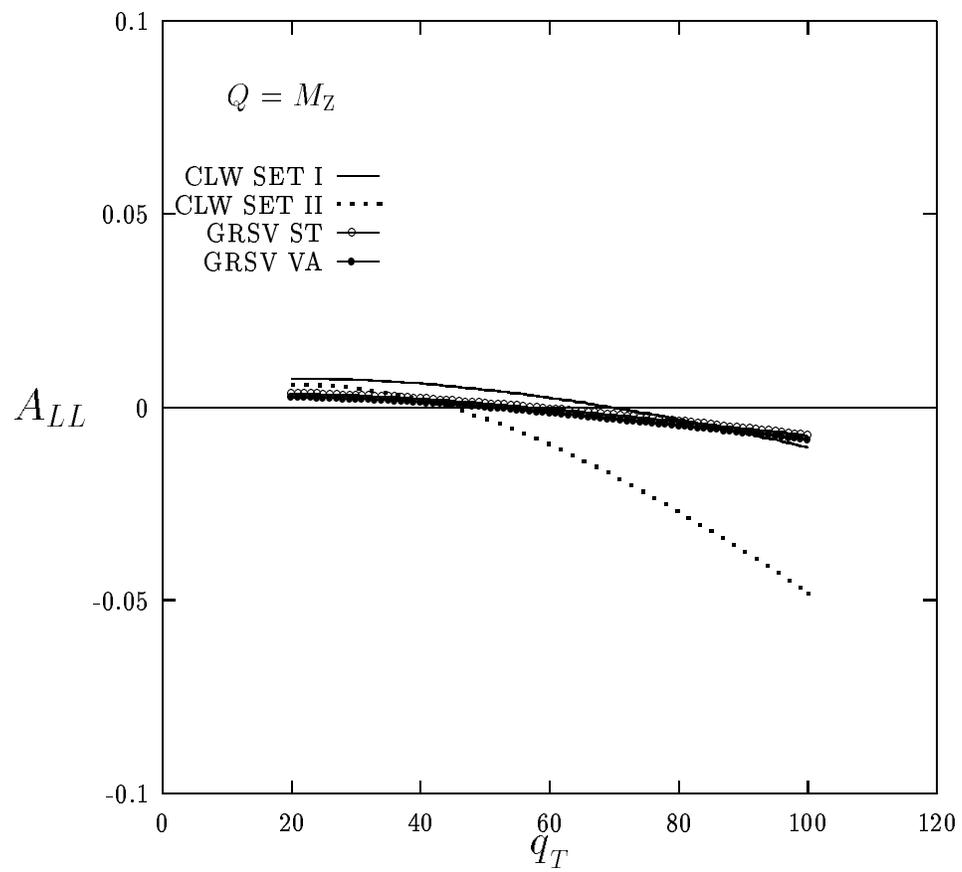
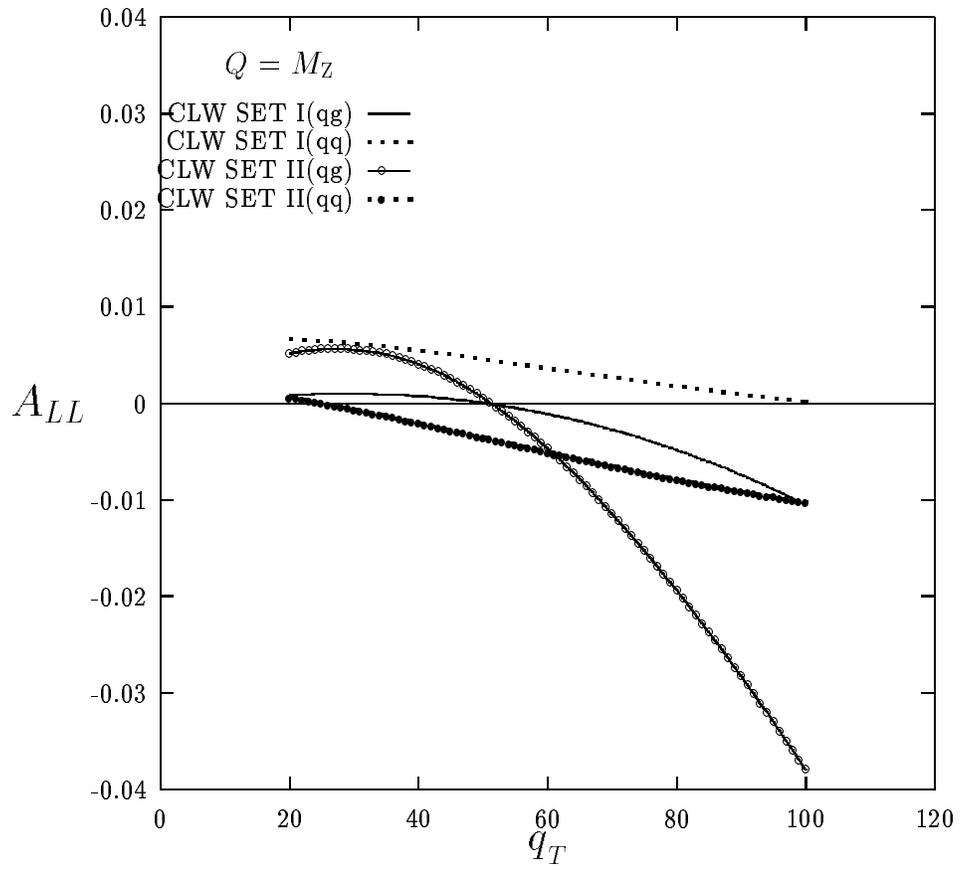
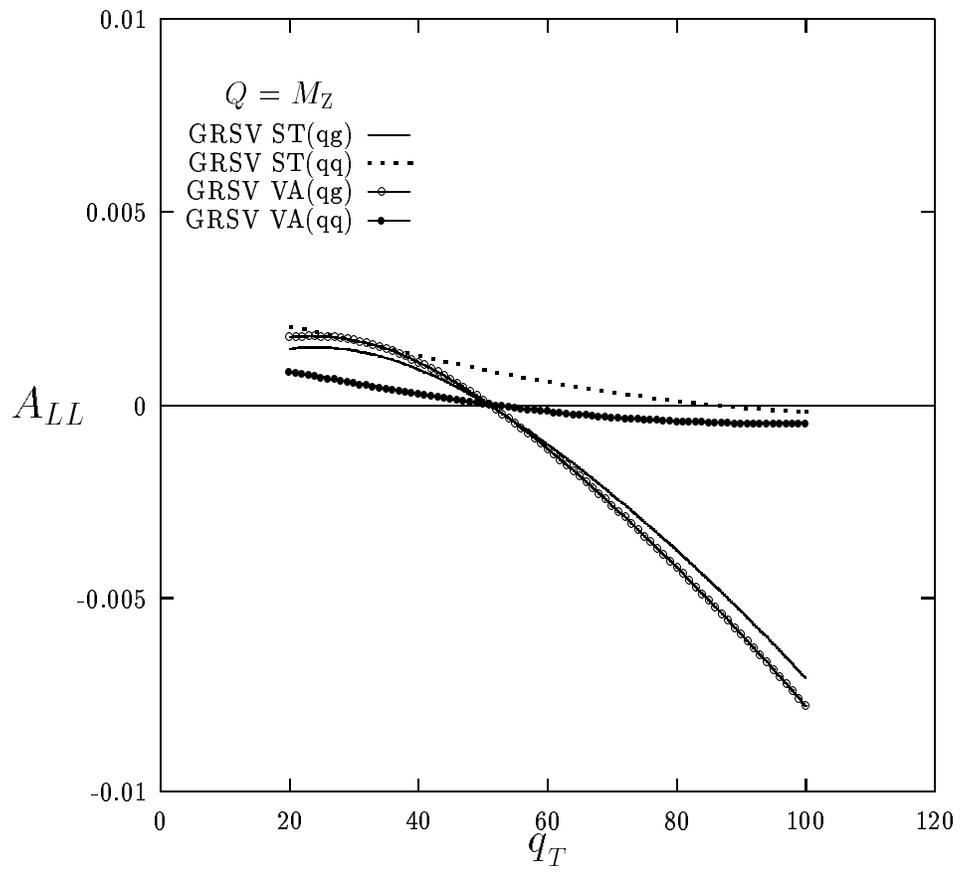
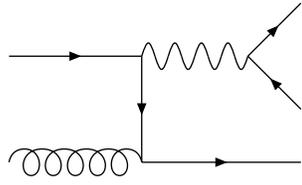
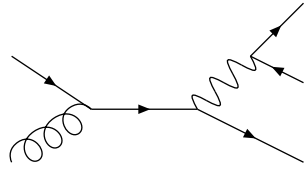
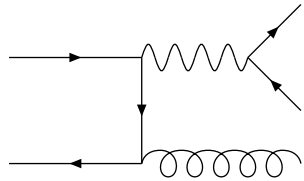
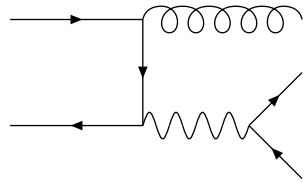
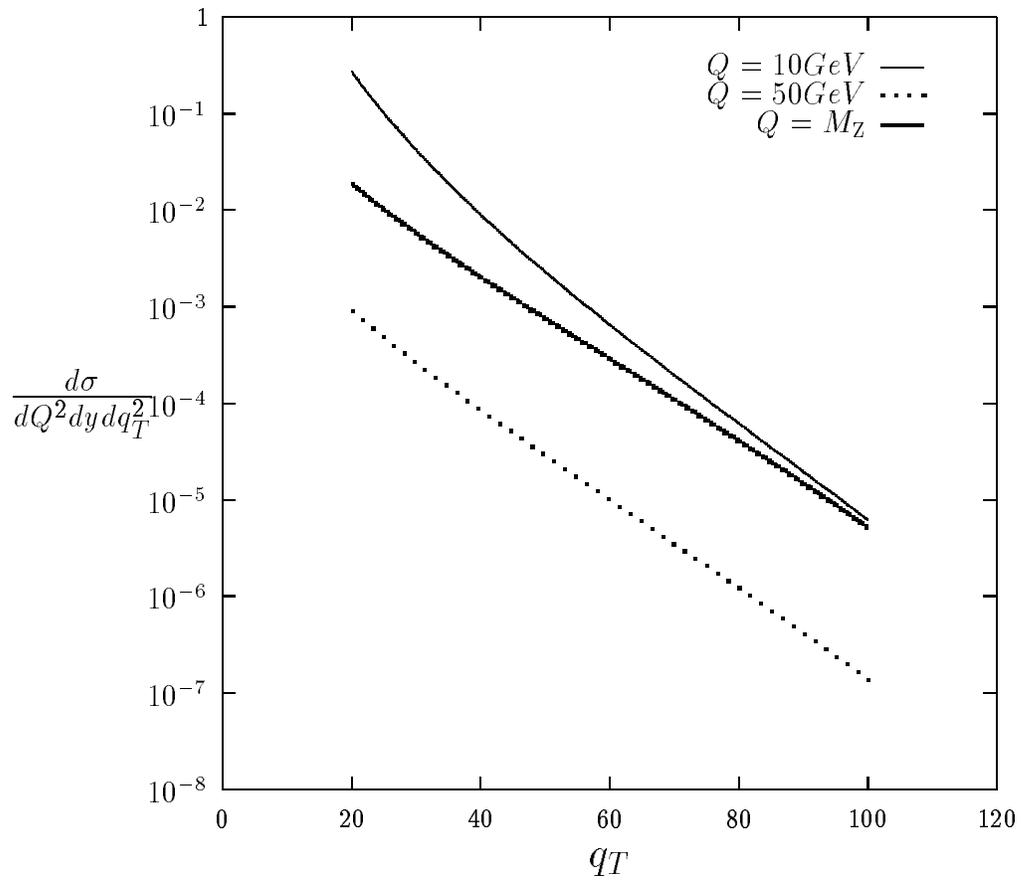
\end{document}